\documentclass{PoS}
\def\red{}
\def\blue{}
\def\et{{\it et al.}}
\def\prd#1{Phys.\ Rev.\ D {\bf #1}}

\title{Update on onium masses with three flavors of dynamical quarks}

\ShortTitle{Update on onium masses with three flavors of dynamical quarks}

\author{\speaker{Steven Gottlieb},  
        L.~Levkova \thanks{Current address: Physics Department, University of
Utah, Salt Lake City, UT 84112, USA}\\ 
	Department of Physics, Indiana University, Bloomington, Indiana 47405, USA\\
        E-mail: \email{sg@indiana.edu}
}
\author{M.~Di Pierro\\
School of Computer Science, Telecommunications and Information Systems,
DePaul University, Chicago, Illinois 60604, USA}
\author{A.~El-Khadra\\
Physics Department, University of Illinois, Urbana, Illinois 61801, USA}
\author{A.S.~Kronfeld, P.B.~Mackenzie, J.~Simone\\
Fermi National Accelerator Laboratory, Batavia, Illinois 60510, USA}
\author{Fermilab Lattice Collaboration and MILC Collaboration}

\abstract{We update results presented at Lattice 2005 on charmonium masses.
New ensembles of gauge configurations with 2+1 flavors of improved
staggered quarks have been analyzed.  Statistics have been increased for
other ensembles.  New results are also available for $P$-wave mesons and for
bottomonium on selected ensembles.
}

\FullConference{XXIVth International Symposium on Lattice Field Theory\\
                July 23-28, 2006\\
                Tucson, Arizona, USA}

\begin{document}

\section{Introduction}
The calculation of the onium spectrum from first principles from lattice QCD
represents a significant goal.  There are a number of states that are
stable to strong decay and far from thresholds (that might make finite size
effects more significant).  It should be possible to accurately calculate
all of their masses.  Further, since there are no valence up and down quarks,
we have to consider only the sea quark mass dependence.
However, one must be careful in
dealing with heavy quarks on the lattice because $aM$ is not small.  
Some early results using dynamical quark configurations were published in
2004 \cite{PRL}.
Using clover type quarks with the Fermilab interpretation \cite{kkm},
we have been calculating the onium spectrum for some time \cite{onium2003}
on MILC gauge configurations \cite{MILCLATS}.  
The HPQCD/UKQCD collaborations have successfully been using NRQCD to treat the 
bottom quark on many of the same ensembles \cite{DAVIES}.
Recently, they have started to use highly improved staggered quarks (HISQ) 
to study charm \cite{hisq}.

At Lattice 2005 \cite{onium2005}, we presented results for lattice spacings
$a \approx 0.18$, 0.12 and 0.09 fm.  These are denoted extra coarse, coarse
and fine, respectively.
This year we have several new ensembles
with a lattice spacing of $\approx 0.15 $ fm, denoted medium coarse 
in the plots.  On these new ensembles, we have
tuned the dynamical strange quark mass closer to its physical value based
upon experience with the earlier ensembles.  At the fine lattice spacing,
we have new results on a more chiral ensemble with $am_l=0.0031$ and
$am_s=0.031$ on a $40^3\times 96$ grid.
All of our results for bottomonium are new.  We also have some new results for
the $\chi_{c2}$.  (See Table~\ref{lats} for details of the ensembles.)


\begin{table}[b]
\begin{center}
\begin{tabular}{|c|c|c|c|c|c|}
\hline
$am_q$ / $am_s$  & \hspace{-1.0mm}$10/g^2$ &$\approx a$ & size & volume & config. \\
\hline
0.0492  / 0.082   & 6.503 & 0.18 & $16^3\times48$ & $(2.8\;{\rm fm})^3$ & 401 \\
0.0328  / 0.082   & 6.485 & 0.18 & $16^3\times48$ & $(2.8\;{\rm fm})^3$ & 331 \\
0.0164  / 0.082   & 6.467 & 0.18 & $16^3\times48$ & $(2.8\;{\rm fm})^3$ & 645 \\
0.0082  / 0.082   & 6.458 & 0.18 & $16^3\times48$ & $(2.8\;{\rm fm})^3$ & 400 \\
\hline
0.0194  / 0.0484   & 6.586 & 0.15 & $16^3\times48$ & $(2.4\;{\rm fm})^3$ & 631 \\
0.0097  / 0.0484   & 6.566 & 0.15 & $20^3\times48$ & $(3.0\;{\rm fm})^3$ & 631 \\
\hline
0.03  / 0.05   & 6.81 & 0.12 &$20^3\times64$ & $(2.4\;{\rm fm})^3$ & 549 \\
0.02  / 0.05   & 6.79 & 0.12 &$20^3\times64$ & $(2.4\;{\rm fm})^3$ & 460 \\
0.01  / 0.05   & 6.76 & 0.12 &$20^3\times64$ & $(2.4\;{\rm fm})^3$ & 593 \\
0.007  / 0.05   & 6.76 & 0.12 &$20^3\times64$ & $(2.4\;{\rm fm})^3$ & 403 \\
0.005  / 0.05   & 6.76 & 0.12 &$24^3\times64$ & $(2.9\;{\rm fm})^3$ &  397 \\
\hline
0.0124  / 0.031   & 7.11 & 0.09 & $28^3\times96$ & $(2.4\;{\rm fm})^3$ & 517 \\
0.0062  / 0.031   & 7.09 & 0.09 & $28^3\times96$ & $(2.4\;{\rm fm})^3$ & 557  \\
0.0031  / 0.031   & 7.08 & 0.09 & $40^3\times96$ & $(3.4\;{\rm fm})^3$ & 504  \\
\hline
\end{tabular}
\caption{Ensembles used in this calculation.}
\label{lats}
\end{center}
\end{table}

To calculate the onium spectrum, we need to find an appropriate value of
the heavy quark hopping parameter $\kappa$.  We do this by studying the
$D_s$ and $B_s$ kinetic masses.  We do this study on one ensemble for each
lattice spacing and use the selected values of $\kappa_c$ and $\kappa_b$
for all the ensembles with that lattice spacing.
Errors in the kinetic masses tend to be large, and so we calculate mass
splittings based on differences in the rest energy.  


\section{Charmonium Results}

\begin{figure}[tb]
  \hfill
    \begin{minipage}[t]{0.45\textwidth}
      \begin{center}
      \includegraphics[width=\textwidth]{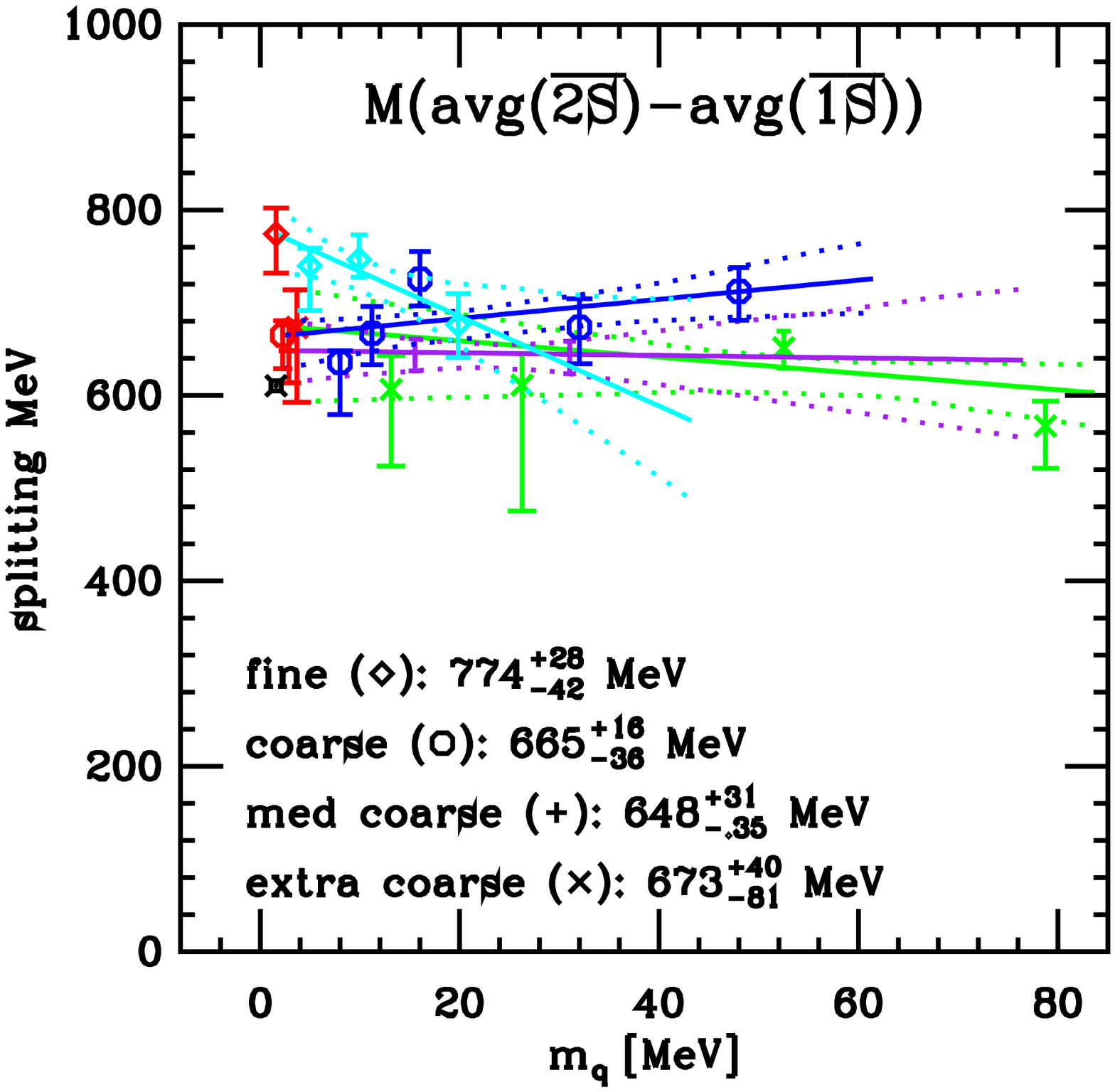}
      \caption{Splitting between the spin-averaged $2S$ and $1S$ states of 
      charmonium.}
      \label{2s1s}
      \end{center}
    \end{minipage}
  \hfill
    \begin{minipage}[t]{0.45\textwidth}
      \begin{center}
       \includegraphics[width=\textwidth]{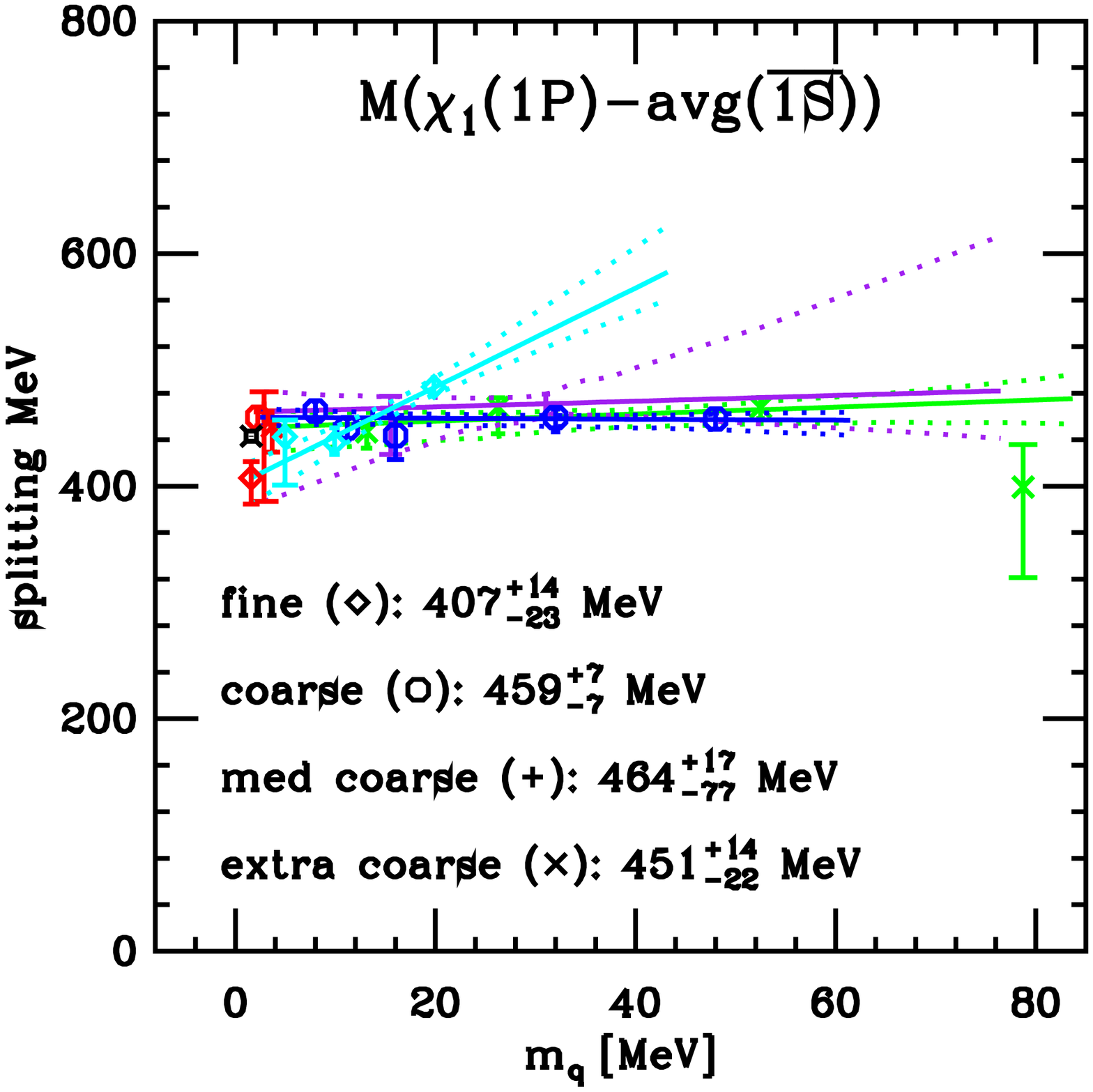}
       \caption{Splitting between the $\chi_{c1}(1P)$ and spin-averaged $1S$ 
		states.}
      \label{chic1P}
      \end{center}
    \end{minipage}
  \hfill
\end{figure}

We will examine a number of splittings in the charmonium spectrum.  Let's
look at the difference between the spin averaged $2S$ and $1S$ states.  In
Fig.~\ref{2s1s}, we plot the splitting in MeV as a function of the  light bare sea
quark mass.  The experimental value is shown as a black fancy cross.  The
points in red are linear extrapolations in quark mass.  We find that at
each lattice spacing (except for the fine lattice) the extrapolated value is
in good agreement with experiment.  On the fine lattice, there is a
considerable slope and the extrapolated value is quite high.  (We note that
as the horizonal axis is the bare quark mass, the slope is not a physical
quantity; it also reflects a change in the mass renormalization as the
lattice spacing changes.)

Next we turn to issues of fine structure and look at the splitting between
the $\chi_{c1}(1P)$ and the spin averaged $1S$ mass.  It would be more natural
to compare with the spin average of all the $1P$ states, but the
$\chi_{c2}$ mass is not yet available on all ensembles, so we compare with
the $1S$ spin average.  
Using the same color scheme and symbols as in the
previous figure, we see in Fig.~\ref{chic1P},
that there is good agreement with experiment
except on the fine ensembles which again exhibit a substantial slope.  In
this case, the two more chiral ensembles are in good agreement with the
experimental value, but the ensemble with the heaviest value of the light
sea quark mass gives too large a value for the splitting, leading to a 
large slope and too small a chiral limit.

\begin{figure}[tbh]
  \hfill
    \begin{minipage}[t]{0.45\textwidth}
      \begin{center}
      \includegraphics[width=\textwidth]{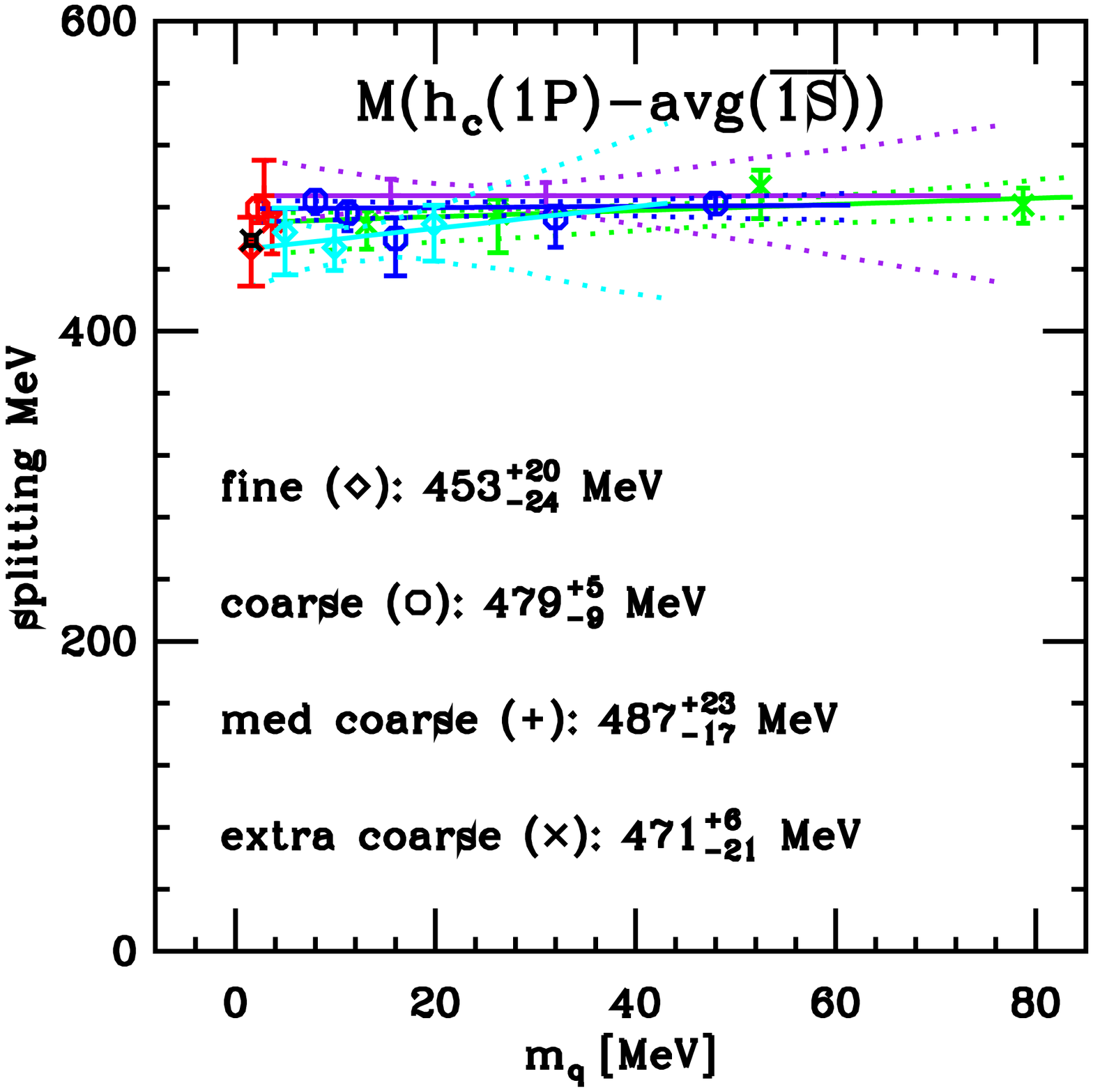}
      \caption{Splitting between the $h_{c}(1P)$ and spin-averaged $1S$ states.}
      \label{hc1P}
      \end{center}
    \end{minipage}
  \hfill
    \begin{minipage}[t]{0.45\textwidth}
      \begin{center}
       \includegraphics[width=\textwidth]{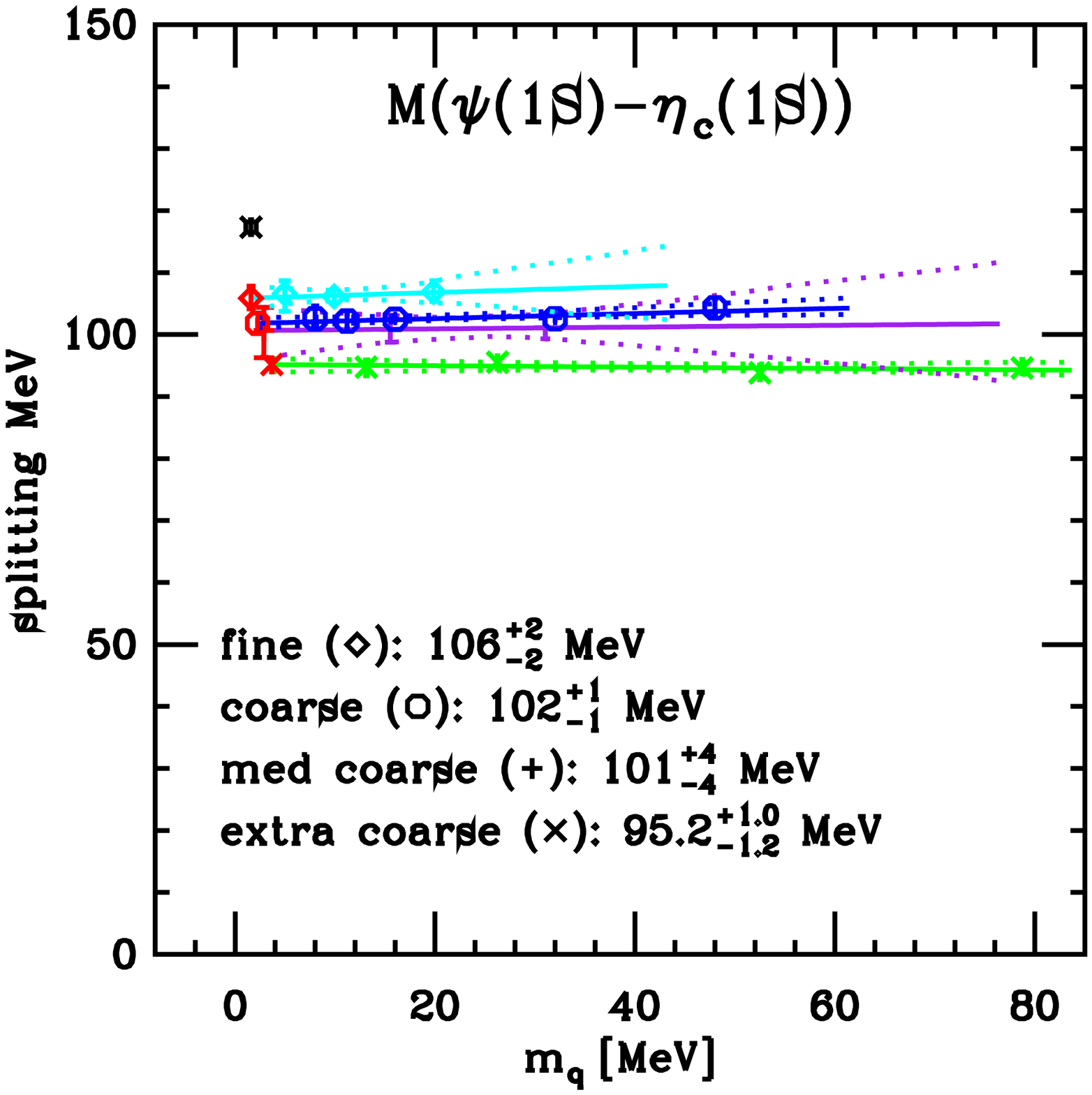}
       \caption{Hyperfine splitting of the $1S$ states.}
       \label{psi-etac}
      \end{center}
    \end{minipage}
  \hfill
\end{figure}

For the $h_c(1P)$ state shown in Fig.~\ref{hc1P}, we find good agreement with
the experimental value on all ensembles.  For the finest lattice spacing,
there is only a modest slope from the chiral extrapolation. 

Turning next to the hyperfine splitting, we look at the $J/\psi(1S)$ --
$\eta_c(1S)$ mass splitting in Fig.~\ref{psi-etac}.  In this case, we find
that the splittings are sytematically small, but that the value is
increasing toward the experimental value as the lattice spacing decreases.
 
\begin{figure}
\begin{center}
\includegraphics[width=0.45\textwidth]{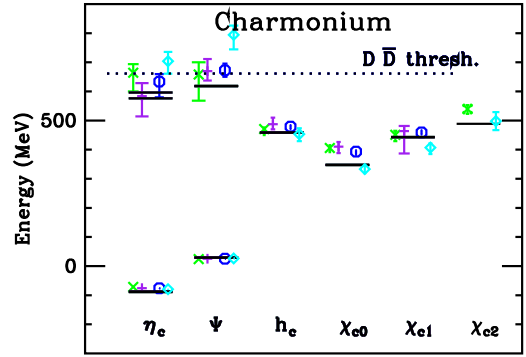}
\end{center}
\caption{Summary of charmonium spectrum.}
\label{charmsummary}
\end{figure}

The $\chi_{c2}(1P)$ has only been studied on two ensembles so far.  We have
new results on one fine ensemble.  In Fig.~\ref{charmsummary}, we summarize
the results for all the states studied.  Except for the $\chi_{c2}(1P)$, we
plot results from our linear chiral extrapolation for each lattice spacing.
For the ground states, if we focus our attention on the diamonds
representing our smallest lattice spacing, we find the most serious
discrepancy between our results and experiment is for the $\chi_{c1}$.  We
have seen that our linear chiral extrapolation may be the culprit here, as the
two more chiral ensembles are in good agreement with the experimental
value.  The $S$ wave first excited states are not that well determined, but
are rather heavy compared to the observed values.  We have seen that on the
finest lattice spacing, the high slope of the chiral extrapolation is
accentuating the difference between our calculation and observations.
Furthermore, the observed states are quite close to the $D\bar D$ threshold,
which makes these states harder to calculate on the lattice without careful
attention to finite volume effects.  Thus, we are not seriously concerned 
about the high masses we are seeing for the 2$S$ states.

\section{Bottomonium Results}

We have new results this year for bottomonium on the three smallest lattice
spacings.  Some important ground states in this system, the $\eta_b$ and
$h_b$, have not been observed, so we shall modify some of the splittings we
display.  It is also interesting to compare our results with those using
NRQCD for the heavy quarks \cite{DAVIES}.  
We expect that our results have larger discretization errors, when comparing
them to the results of Ref.~\cite{DAVIES} at a fixed lattice spacing, because
the NRQCD action used in that work is more improved.

\begin{figure}[tb]
  \hfill
    \begin{minipage}[t]{0.45\textwidth}
      \begin{center}
      \includegraphics[width=\textwidth]{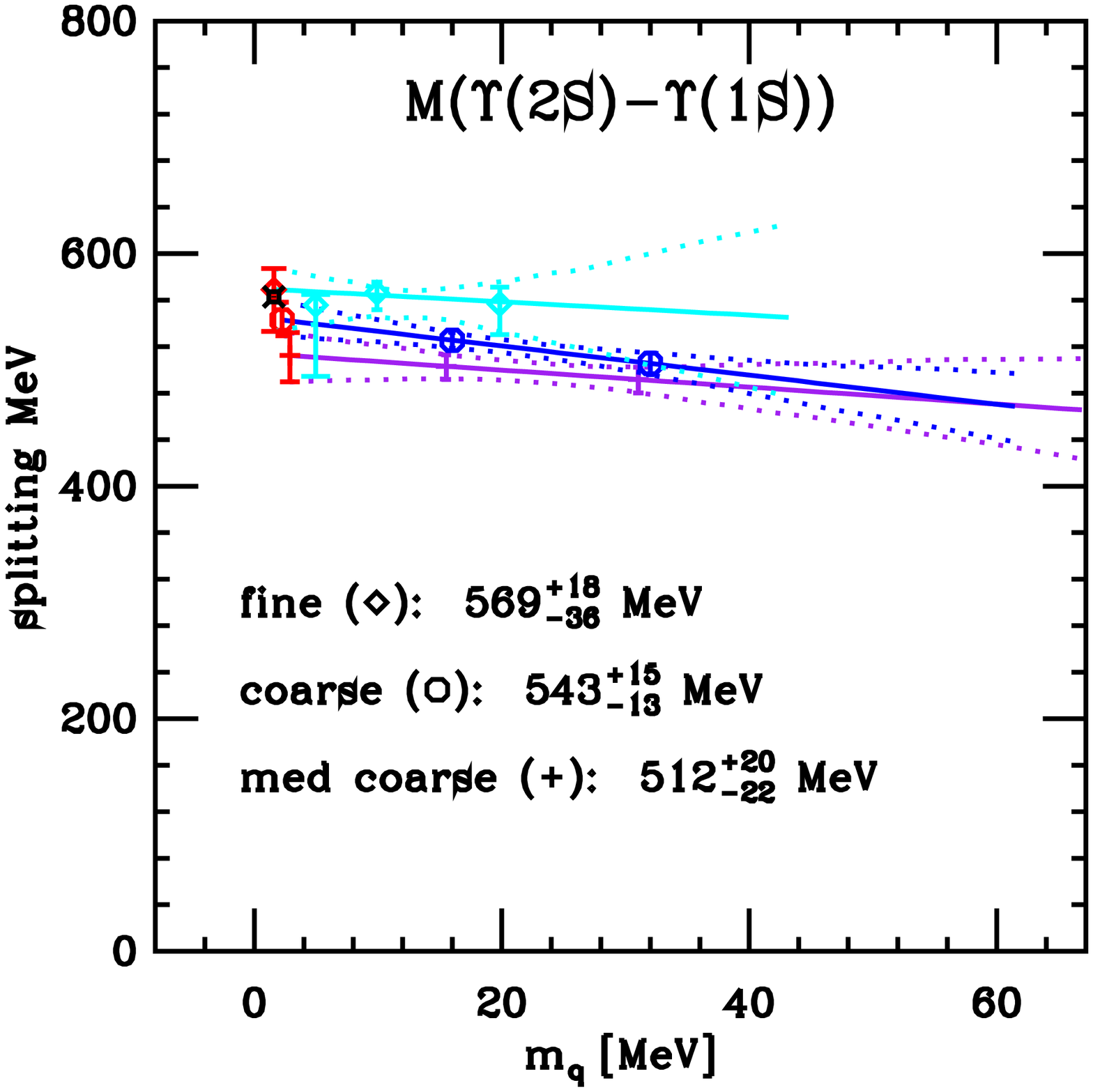}
       \caption{Chiral extrapolation of the $\Upsilon(2S)$-$\Upsilon(1S)$ 
		splitting.}
       \label{upsilon2s1s}
      \end{center}
    \end{minipage}
  \hfill
    \begin{minipage}[t]{0.45\textwidth}
      \begin{center}
       \includegraphics[width=\textwidth]{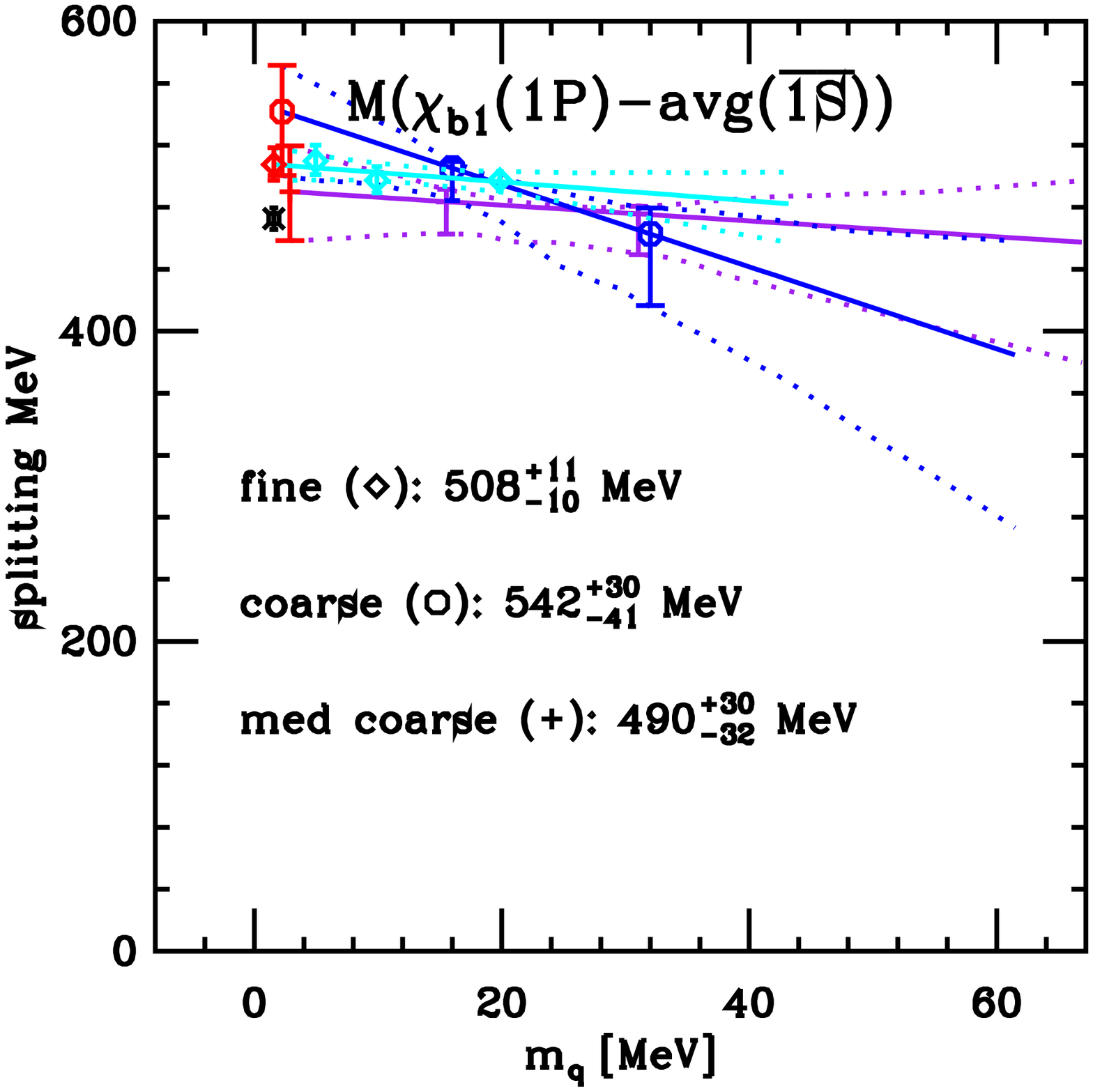}
       \caption{Chiral extrapolation of the $\chi_{b1}(1P)$ -- spin averaged 
		$1S$ splitting.}
       \label{chi_b1-1savg}
      \end{center}
    \end{minipage}
  \hfill
\end{figure}

In Fig.~\ref{upsilon2s1s}, we plot the splitting between the $\Upsilon(2S)$
and $\Upsilon(1S)$ masses.  On the fine ensembles, the result is in good 
agreement with experiment.  With larger lattice spacings, the splitting is a bit
low.  We look only at the $\Upsilon$ level since the $\eta_b$ masses are
not well measured.  In the bottomonium system, the $2S$ and $3S$ states are 
both below the $B\bar B$ threshold.  Thus, possible finite size effects from a
nearby threshold are not an issue, and we should get the $2S$ levels right.

In considering the fine structure of the bottomonium spectrum we are faced
with two issues.  First,
the $h_b$ has not been observed so we cannot compare our results with the
experimental value (but we can make a prediction).  
Second, we have no results for $\chi_{b2}$ yet, so we
can't compute a spin average of the $P$ states.  So,
we consider the $\chi_{b1}$, in the middle of the three ${}^3P_J$ states.
We compare it to the spin averaged $1S$ states assuming our
$\eta_b$ mass is correct.
The result shown in
Fig.~\ref{chi_b1-1savg} has the splitting about 35 MeV too large on the
fine lattices.

\begin{figure}[tb]
  \hfill
    \begin{minipage}[t]{0.45\textwidth}
      \begin{center}
      \includegraphics[width=\textwidth]{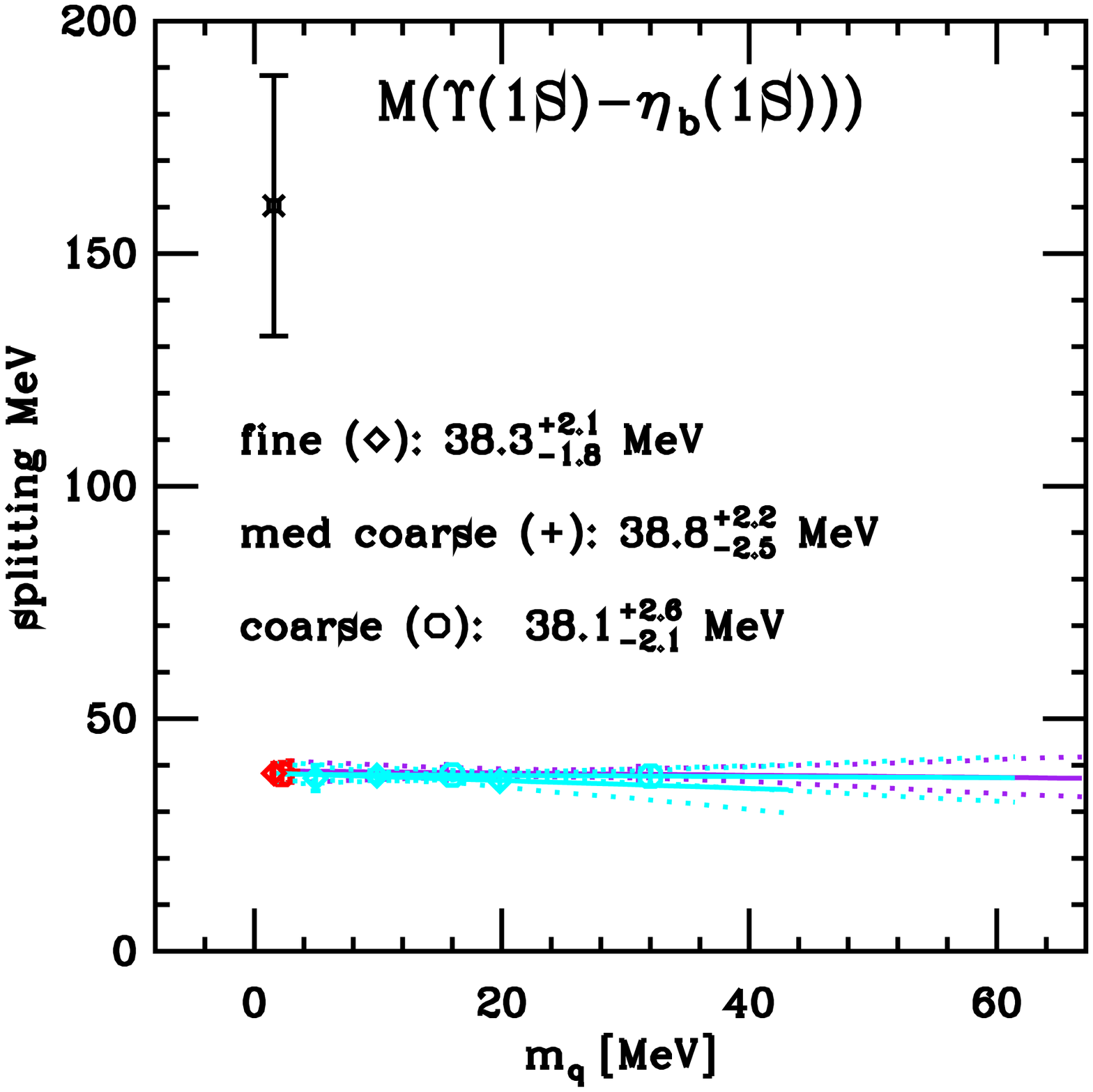}
       \caption{Hyperfine splitting of the bottomonium $1S$ states.}
       \label{hyperfine1s}
      \end{center}
    \end{minipage}
  \hfill
    \begin{minipage}[t]{0.45\textwidth}
      \begin{center}
       \includegraphics[width=\textwidth]{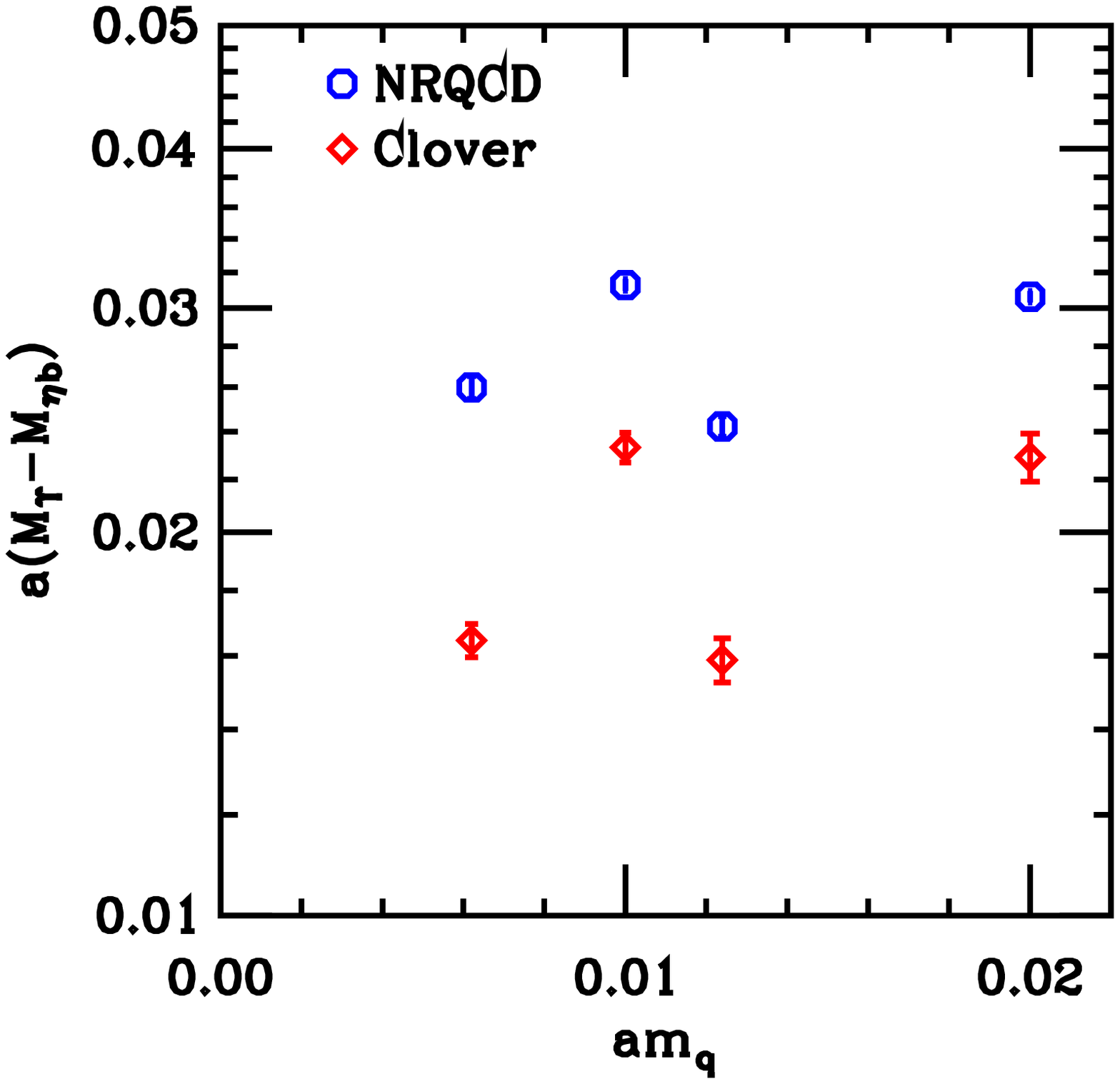}
       \caption{Comparison of HPQCD's hyperfine splitting using NRQCD with 
		this work.}
       \label{hpqcdhyperfine}
      \end{center}
    \end{minipage}
  \hfill
\end{figure}

The hyperfine splitting in the $1S$ states is shown in Fig.~\ref{hyperfine1s}.
We find a value of about 38 MeV for this splitting for all $a$.  The
experimental result plotted comes from the PDG and is based on the
observation of a single $\eta_b$.  It should be noted that a preliminary
result from CDF with higher statistics had a splitting of only 15 MeV.
With this factor of 10 difference between the experimental results, it is
hard to reach any definite conclusion about how accurate our hyperfine
splitting is for the $1S$ states.  It turns out, however, that on four
ensembles we can directly compare our calculation with the HPQCD results.
In Fig.~\ref{hpqcdhyperfine}, we compare the splitting in lattice units.
The results are plotted on a semilogarithmic scale so that equal vertical
differences represent equal fractional differences in the splittings.  The
upper two octagons and diamonds are coarse lattice results.  The lower two
octagons and diamonds are on fine ensembles.  Points with the same bare
quark mass are calculated on the same ensembles, but the HPQCD
collaboration did not analyze every available configuration, so the actual set
of configurations selected for analysis differs.  We see that the fraction
difference in the hyperfine splittings is larger at the smaller lattice
spacing and that the NRQCD hyperfine splitting is larger than that for clover.
The difference is not unexpected, because the leading error for this splitting
with the clover action is of order $mv^6$, whereas in Ref.~\cite{DAVIES} it is
of order $mv^8$.

\begin{figure}
\begin{center}
\includegraphics[width=0.45\textwidth]{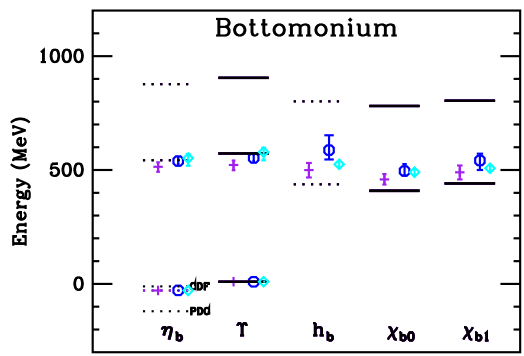}
\end{center}
\caption{Summary of the bottomonium spectrum for ensembles with 
$a\approx 0.15$, 0.12 and 0.09 fm.}
\label{bsummary}
\end{figure}

In Fig.~\ref{bsummary},
we summarize the spectrum of observed states using the spin-averged $1S$ states
(assuming a 38 MeV $\Upsilon$--$\eta_b$ splitting)
to set the additive constant needed to go from splittings to masses.
Solid lines on
the plot are experimentally determined masses.  Dashed lines on
the plot denote unobserved (or poorly observed) states.  
In the case of the $\eta_b$ we use
dashed lines and mark them CDF and PDG as discussed above.  We find good
agreement with experiment for the $\Upsilon(2S)$.  The $\eta_b(2S)$ is in
good agreement with theoretical expectations.  Our masses for the $P$ wave
states $h_b$, $\chi_{b0}$ and $\chi_{b1}$ are too large.

\section{Conclusions and Outlook}

We observe only a mild dependence on the sea quark mass for most of the
quantities studied here.  In particular, the hyperfine splittings and the
$2S$-$1S$ splittings appear to be essentially independent of the sea quark
mass.  Hence a linear extrapolation seems adequate.  For the $P$-wave states,
we do observe a sea quark mass dependence at the 10\% level on some
ensembles.

There are several positive features of the charmonium spectrum.  The $h_c$
and $\chi_{c1}$ masses look quite good (although the latter is driven low
by the chiral extrapolation on the fine ensembles).  At the smallest
lattice spacing studied so far, the $P$ wave splitting looks good 
(but see previous sentence).  The hyperfine splitting is
too small but improving as the lattice spacing decreases.  
However, the $2S$ states
are not accurately calculated.  For the bottomonium spectrum, the excited
state splitting looks good.  It is not yet possible to test the hyperfine
splitting, but our splitting is smaller than that coming from NRQCD
calculations.  Our $P$ wave  states seem too heavy.

In the coming year, we expect to increase our statistics on a number of
existing ensembles.  In addition, MILC is generating ensembles with a
lattice spacing of 0.06 fm and has plans to reduce the lattice spacing to
0.045.  There is also some chance that there will be some ensembles with
$a=0.105$ fm.
Currently, we are using an automatic criterion for picking the best fit.  We
need to consider alternative methods and whether picking a single fit
properly reflects the systematic errors.

The $P$ wave splitting in the bottomonium spectrum does not seem to be in
agreement with experiment.  
It will be interesting to see if the bottomonium spectrum improves as
we reduce the lattice spacing (and $am_b$).  
Kronfeld and Oktay \cite{Oktay} have been developing a highly improved 
clover quark action that we hope to use in the future.

\end{document}